\documentstyle{article}
\begin{document}

\centerline {\bf  COMBINATORIAL EXPRESSION FOR UNIVERSAL VASSILIEV LINK
INVARIANT}
$$  $$
\centerline {\bf    SERGEY PIUNIKHIN}
\centerline { }
\centerline {\bf   Harvard University}
\centerline { }
\centerline  {February 18 , 1993}

$$  $$

 \centerline {\bf         ABSTRACT}
\centerline { }

     The most general R-matrix type state sum model for link
invariants is constructed. It contains in itself all
R-matrix invariants and is a generating function for "universal"
Vassiliev link invariants. This expression is more simple than
Kontsevich's expression for the same quantity,  because it   is
defined combinatorially
and does not contain any integrals, except for an expression for "the universal
Drinfeld's associator".
$$ $$
\centerline {\bf     1. INTRODUCTION}
\centerline { }
     Vassiliev knot invariants were invented [Va1], [Va2] in
attempt to construct some natural basis in the space
$H^0$ (Imbeddings: $S^1 \rightarrow {\bf R}^3$ )  of all knot
invariants,  using natural
stratification of discriminant set of nonimbeddings:
$S^1 \rightarrow {\bf R}^3$ and a kind of infinite-dimensional Alexander
duality. Although
the question whether Vassiliev knot invariants can distinguish
any two knots is still opened,  this language is the most
appropriate in studying classical knot and link invariants.
All known knot and link invariants: Alexander polynomial [Ro],
[Co],  Jones polynomial [Jo1],  Kauffman polynomial [Ka1], [Ka2],
HOMFLY polynomial [HOMFLY] and all their generalizations [Tu1];[Tu2];[Re1];[RT]
and Milnor's  $\mu $-invariants [Mi1], [Mi2],  which generalize Gauss
linking number -can be incorporated into this scheme (see [BL],
[Li1],  [Li2], [BN5]).

     The space of Vassiliev knot invariants of fixed order  n
(factored through the space of invariants of order n-1) can be
described pure combinatorially as a space of functions on the set
of assignments of  n  pairs of points on $\ S^1$   subject to certain
linear relations. These relations were first written in [BL],  the
fact,  that this set of relations is complete was proved by
Kontsevich [Ko1],  using explicit integral presentation of
"universal" Vassiliev invariant  $I_n$  of  order  n,  which takes
values in a certain linear space $F_n$ generated by "Vassiliev
diagrams". This space has also very nice description in terms of
Feynman diagrams of perturbative Chern-Simons theory [BN1], [BN2],
[BN3], [Ko1]; the graded linear space  $F = \oplus_n F_n$  admits Hopf
algebra structure [Ko1], [BN3] (Kontsevich Hopf algebra). The
space of primitive elements in this Hopf algebra is generated by
connected Feynman diagrams [Pi2], [Ko2], [BN1].

	 The generating
function
$I = \sum_{n=0}^{\infty}  h^n I_n$  of "universal order n" - Vassiliev
invariants gives us the "universal"  F-valued Vassiliev invariant
(here  h  is formal parameter),  where  $I_n$(K) is certain  n-fold
integral over the knot  K (Kontsevich integral [Ko1]). At the moment
nobody is able to calculate explicitly  I(K)  for any non-trivial
knot  K.

	The aim of the present paper is to give more simple
expression for this quantity,  which can be calculated explicitly in
all orders in h. , if one can calculate the universal Drinfeld's "associator"
[Dr1],[Dr2] . This expression models state sum expression
[Re1], [Tu1], [Jo2] for knot polynomials   $P_{g, V} (q^{\pm 1})$ (here $ q =
e^h$ )
constructed from (semi)simple Lie algebra  g  and its irreducible
representation  V,  after forgetting about  g  and  V . Connection
between  $P_{g, V}(q^{\pm 1})$  and Vassiliev knot invariants was found in
[Li1] : If $P_{g, V}$(h)= $\sum_{n=0}^{infty} P_{g, V, n} h^n$   than $P_{g, V,
n}$        is Vassiliev invariant
of order  n. Explicit state sum expression for  $P_{g, V, n}\in  F_n$  was
deduced in [Pi1],  using results of [BN1],  where the same quantity
was derived  from perturbative Chern-Simons theory [BN3], [AS], [GMM],
[Ko3]. The generating function  $P_{g, V}$(h)= $\sum_{n=0}^{infty} P_{g, V, n}
h^n$   can be
explicitly calculated as a state sum expression in terms of knot
diagram [Re1], [Tu1], [Jo2].

     The question is,  whether it is possible to forget about Lie
algebra g  and representation  V  and to write the "universal" state sum
expression $P= \sum_{n=0}^{infty} h^n P_n$   with values in Kontsevich Hopf
algebra  F
and  then,  in order to obtain  $P_{g, V}$  ,  apply Bar-Natan rules to
$P= \sum_{n=0}^{infty} h^n P_n$ ;$P \in F$ to obtain  C-valued invariant $P_{g,
V, n}$ . Half of
this program  (forgetting  about  V  but not  about  g) was
accomplished by Reshetikhin [Re2] where he constructed knot
invariant $P_g \in Z(U(g))$ belonging to the center of quantum
universal enveloping algebra U (g). Here  $P_{g, V}  = Tr_V (P )$. This
construction also works for an arbitrary quasitriangular Hopf
algebra [Dr3] instead of U(g)  [Re2] and for arbitrary
quasitriangular quasi-Hopf algebra [Dr1]. The next stage is to
forget about this quasi-Hopf algebra. There are two ways to do this.
The first one is to forget about  U(g) "in the integral expression
for monodromy of Knizhnik-Zamolodchikov equation [KZ], [K1], [TK].
This is the  way  how  Kontsevich  obtained  his  famous  integral
expression for I  [Ko1].
     The second way is to use "local" braid group representation
$B_m \rightarrow (U(g)\otimes C[h])$   using Drinfeld's  R-matrix
$R = e^{ht \over 2} \in (U(g))\otimes C[h])^{\otimes 2}$    and "associator"
$\Phi \in (U(g))\otimes C[h])^{\otimes 3}$
[Dr1], [Dr2] and then to forget about  U(g). This way will give us the
desired expression.

     The parer is organized as follows:

     In section 2 some preliminary definitions are given and "the universal
prounipotent"
braid group representation is
constructed.

     In section 3 F-valued "Markov trace" in this representation
is constructed and the fact that it is a generating function for
"universal" Vassiliev link invariants is proved. Multiplicative property of
this "universal"
invariant with respect to connected sums is proved . Generalization
for string link invariants is also given.

     In  section 4 some discussions are made.
$$ $$
\centerline {\bf     2. PRELIMINARIES}
\centerline { }
     Following D.Bar-Natan,  we shall call BN-diagram several directed
circles (called Wilson loops) and certain finite number of dashed
lines (called gauge propagators). The propagators and  Wilson
lines are allowed to meet in two types of vertices: one type (called
$R^2 g$-vertices) in which a propagator ends on one of the Wilson loops;
and another (called $g^3$ -vertices) connecting three propagators. We
assume,  that one of two possible orders of gauge propagators meeting
in any  $g^3$ -vertex is specified.
     Each BN-diagram can be uniquely presented by isotopy type of
plane diagrams such as in fig.1.

\setlength{\unitlength}{0.0125in}%
\begin{picture}(289,104)(83,623)
\thicklines
\put(320,675){\oval(104,104)}
\put(135,675){\oval(104,104)}
\multiput(135,655)(-7.69231,0.00000){7}{\line(-1, 0){  3.846}}
\multiput(135,655)(7.69231,0.00000){7}{\line( 1, 0){  3.846}}
\multiput(135,725)(0.00000,-8.23529){9}{\line( 0,-1){  4.118}}
\multiput(320,725)(0.00000,-8.00000){13}{\line( 0,-1){  4.000}}
\multiput(185,680)(8.09524,0.00000){11}{\line( 1, 0){  4.048}}
\end{picture}

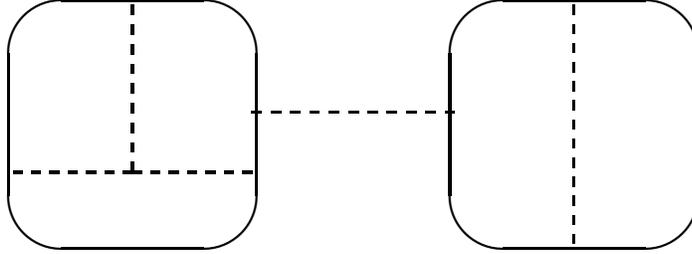
\centerline {\bf fig.1 }
\centerline { }
Here,  as usual,  we assume,  that all four-valent vertices are
fake,  and the counterclockwise order in each  $g^3$ -vertex is
fixed. For instance,  the orders of propagators on graphs
in fig.2a  and fig.2b are different.

\setlength{\unitlength}{0.0125in}%
\begin{picture}(440,85)(60,625)
\put(320,633){\makebox(0,0)[lb]{\raisebox{0pt}[0pt][0pt]{\elvrm fig.2b}}}
\put( 70,628){\makebox(0,0)[lb]{\raisebox{0pt}[0pt][0pt]{\elvrm fig. 2a}}}
\thicklines
\multiput(130,680)(29.72317,-24.76931){2}{\line( 6,-5){ 11.097}}
\multiput(130,680)(32.98150,21.98766){2}{\line( 3, 2){ 12.019}}
\multiput(130,680)(-8.23529,0.00000){9}{\line(-1, 0){  4.118}}
\multiput(360,675)(21.54025,-10.77012){3}{\line( 2,-1){ 12.920}}
\multiput(415,645)(24.07077,16.04718){4}{\line( 3, 2){ 12.019}}
\multiput(355,675)(25.62194,10.24877){3}{\line( 5, 2){ 13.411}}
\multiput(420,700)(18.09250,-21.71100){4}{\line( 5,-6){  9.247}}
\multiput(360,675)(-8.23529,0.00000){9}{\line(-1, 0){  4.118}}
\end{picture}

Let  K  be some ring ${\bf Z} \i K \i {\bf C}$.
\proclaim Definition.

A function  C : $(BN-diagrams)\rightarrow K$ is called weight system
if two axioms are satisfied:
$$ C(I) = C(H) - C(X)
  \eqno       (2.1)$$
    where I, H, X,  are BN-diagrams,  identical everywhere except some
    small ball,  where they  look as in fig.3

\setlength{\unitlength}{0.0125in}%
\begin{picture}(403,160)(17,565)
\thicklines
\multiput( 92,715)(0.00000,-7.74194){16}{\line( 0,-1){  3.871}}
\multiput( 17,715)(7.83784,0.00000){19}{\line( 1, 0){  3.919}}
\multiput( 22,595)(8.10811,0.00000){19}{\line( 1, 0){  4.054}}
\put( 72,565){\makebox(0,0)[lb]{\raisebox{0pt}[0pt][0pt]{\elvrm I}}}
\multiput(200,640)(7.82609,0.00000){12}{\line( 1, 0){  3.913}}
\put(285,710){\makebox(0.4444,0.6667){\tenrm .}}
\multiput(290,710)(0.00000,-8.00000){18}{\line( 0,-1){  4.000}}
\multiput(200,575)(0.00000,8.00000){18}{\line( 0, 1){  4.000}}
\put(350,605){\makebox(0.4444,0.6667){\tenrm .}}
\multiput(355,605)(7.27273,0.00000){6}{\line( 1, 0){  3.636}}
\multiput(325,570)(14.34963,23.91605){7}{\line( 3, 5){  7.432}}
\multiput(420,565)(-14.49670,24.16117){7}{\line(-3, 5){  7.432}}
\put(240,565){\makebox(0,0)[lb]{\raisebox{0pt}[0pt][0pt]{\elvrm H}}}
\put(360,565){\makebox(0,0)[lb]{\raisebox{0pt}[0pt][0pt]{\elvrm X}}}
\end{picture}

\centerline {\bf fig.3 }

$$   C(S) = C(T) - C(U)
\eqno         (2.2)$$    where  S, T, U are BN-diagrams,  identical everywhere
    except some small ball,  where they look as in fig.4

\setlength{\unitlength}{0.0125in}%
\begin{picture}(510,240)(10,450)
\thicklines
\multiput( 85,590)(0.00000,-8.09524){11}{\line( 0,-1){  4.048}}
\multiput( 15,685)(15.53333,-20.71111){5}{\line( 3,-4){  8.667}}
\multiput( 85,590)(17.08163,28.46939){4}{\line( 3, 5){  7.432}}
\put( 10,505){\vector( 1, 0){155}}
\multiput(245,505)(-6.52397,26.09587){7}{\line(-1, 4){  3.503}}
\multiput(265,505)(5.36120,26.80601){7}{\line( 1, 5){  2.833}}
\put(185,505){\vector( 1, 0){150}}
\multiput(420,510)(6.52397,26.09587){7}{\line( 1, 4){  3.503}}
\multiput(445,510)(-10.08293,25.20732){7}{\line(-2, 5){  5.365}}
\put(355,510){\vector( 1, 0){165}}
\put(225,450){\makebox(0,0)[lb]{\raisebox{0pt}[0pt][0pt]{\elvrm fig. 4}}}
\put( 70,495){\makebox(0,0)[lb]{\raisebox{0pt}[0pt][0pt]{\elvrm S}}}
\put(250,495){\makebox(0,0)[lb]{\raisebox{0pt}[0pt][0pt]{\elvrm T}}}
\put(425,500){\makebox(0,0)[lb]{\raisebox{0pt}[0pt][0pt]{\elvrm U}}}
\end{picture}

\centerline {  }

\proclaim  Definition.

Following Vassiliev [Va1], [Va2] and Birman-Lin [BL] we shall call
BN-diagram with 2n $R^2 g$-vertices and without $g^3$ -vertices Vassiliev
[n]-diagram and BN-diagram with 2n-2  $R^2 g$-vertices and with one
$g^3$ -vertex Vassiliev $<n>$-diagram .
     Let  D   be  $<n>$-diagram. Let us define [n]-diagrams
$D_{1+}$  ,  $D_{1-}$    as [n] diagrams,  obtained from   D   by local
procedure
shown in fig.5
\setlength{\unitlength}{0.0125in}%
\begin{picture}(378,121)(5,574)
\thicklines
\put(329,636){\oval(108,108)}
\multiput(289,676)(15.16986,-25.28311){4}{\line( 3,-5){  7.432}}
\multiput(319,581)(15.16986,25.28311){4}{\line( 3, 5){  7.432}}
\put(320,649){\makebox(0,0)[lb]{\raisebox{0pt}[0pt][0pt]{\elvrm D1-}}}
\put(325,574){\makebox(0,0)[lb]{\raisebox{0pt}[0pt][0pt]{\elvrm z1}}}
\put(365,674){\makebox(0,0)[lb]{\raisebox{0pt}[0pt][0pt]{\elvrm z3}}}
\put(279,676){\makebox(0,0)[lb]{\raisebox{0pt}[0pt][0pt]{\elvrm z2}}}
\put(199,642){\oval(106,106)}
\multiput(154,672)(11.18008,-22.36017){4}{\line( 1,-2){  6.460}}
\multiput(204,592)(9.59112,23.97781){4}{\line( 2, 5){  5.365}}
\put(185,655){\makebox(0,0)[lb]{\raisebox{0pt}[0pt][0pt]{\elvrm D1+}}}
\put(195,575){\makebox(0,0)[lb]{\raisebox{0pt}[0pt][0pt]{\elvrm z1}}}
\put(145,675){\makebox(0,0)[lb]{\raisebox{0pt}[0pt][0pt]{\elvrm z2}}}
\put(239,677){\makebox(0,0)[lb]{\raisebox{0pt}[0pt][0pt]{\elvrm z3}}}
\put( 64,641){\oval(108,108)}
\multiput( 65,585)(0.00000,8.09524){11}{\line( 0, 1){  4.048}}
\multiput( 15,670)(8.26087,0.00000){12}{\line( 1, 0){  4.130}}
\put( 55,578){\makebox(0,0)[lb]{\raisebox{0pt}[0pt][0pt]{\elvrm z1}}}
\put(  5,673){\makebox(0,0)[lb]{\raisebox{0pt}[0pt][0pt]{\elvrm z2}}}
\put(110,668){\makebox(0,0)[lb]{\raisebox{0pt}[0pt][0pt]{\elvrm z3}}}
\put( 40,628){\makebox(0,0)[lb]{\raisebox{0pt}[0pt][0pt]{\elvrm D}}}
\end{picture}

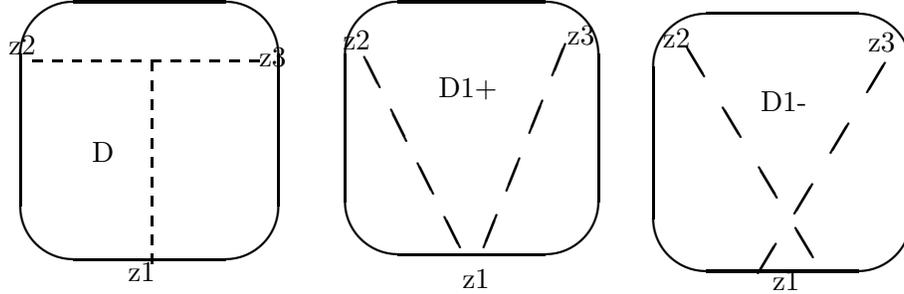
\centerline {\bf fig.5 }

    (The [n]-diagrams  $D_{2+}$  ,  $D_{2-}$ ,  $D_{3+}$  ,  $D_{3-}$   are
defined by
   changing $z_1$   on  $z_2$   and  $z_3$   respectively).
\proclaim  Definition.  [BL], [Va1], [Va2], [Ko1].
     Let  $D^s (s \in \bf N)$    be free  K-module,  generated by the set of
s-Wilson-loop-
Vassiliev [n]-diagrams; let  $F^s$   be its factor-module,  generated
by relations
          $$D_{1+} - D_{1-} = D_{2+} - D_{2-}
\eqno    (2.3)$$
(D   runs over  $<n>$-diagrams).
Let us denote  $F_0^s$ = K;  $F^s = \oplus_n F_n^s$  ,  and let us identify  $1
\in  K = F_0^s$
with (the unique) s-Wilson-loop Vassiliev [0]-diagram.

\proclaim   Theorem 2.1.  [Ko1], [Ar], [BN1]
     K-module $F_n^s$   is isomorphic to factor of free module  $D_n^s$
    generated by  (n+s)-loop BN-diagrams by relations
              $$ I = H - X          \eqno            (2.4)$$
    where I, H, X,  are BN-diagrams,  identical everywhere except
    some small ball,  where they  look as in fig.3 and
              $$S = T - U   \eqno           (2.5)$$
    where  S, T, U are closed diagrams,  identical everywhere
    except some small ball,  where they look as in fig.4 .

     In fact $F^1$ can be provided by the structure of graded Hopf
algebra [Ko1] and we shall call it Kontsevich Hopf algebra. If it will not lead
to confusion, we'll omit the superscript  1  and write  $F = F_1$.

	Kontsevich Hopf algebra $F$ acts  on $F^s$ (taking connected sum along Wilson
loop) in  s  different ways commuting with each other [BN5] , thus we have
graded action of
$F^{\otimes s}$ on $F^s$.

     Let  $A = \oplus_n A_n$   be factor-algebra of Kontsevich algebra by
the ideal,  generated  $F_1$  (this K-module is has rank one and is
generated by the single Vassiliev [1]-diagram. Let us denote this
diagram by  $t \in F_1$ ). Since the element  t  is primitive,  A  is
also Hopf algebra.
     It well-known [Ko1] that the space $A_n^*$   dual to $A_n$ is
canonically isomorphic to the space  $V_n$  of Vassiliev
framed knot invariants of order  n  factored by the space
$V_{n-1}$. The map $V_n/V_{n-1} \rightarrow A_n^*$   is evaluation of knot
invariant
on singular embeddings with  n  double points [Va1], [Va2], [BL]
which gives linear function  $V_n/V_{n-1} \otimes A_n \rightarrow \bf C$ . The
inverse
map  $I_n : A_n^* \rightarrow V_n \rightarrow V_n/V_{n-1}$ was first
constructed in [Ko1]
and is called "Kontsevich integral". The aim of this paper is to
construct (formally another) inverse map $J_n : A_n^* \rightarrow  V_n$ which
does not contain any integrals and which has simple combinatorial
expression.

\proclaim     Definition.   Let  $X^m$  ($m \in \bf N$) be graded completion of
Lie algebra   $\oplus_n X_n^m$ ,  with generators  $t^{ij}$  (i$<$j)  of degree
1 and relations
   $$ [t^{ij}  ; t^{ik} + t^{jk}  ] = 0 \eqno     (2.6) $$
This algebra is dual to Sulivan minimal model of "configuration space" ($\bf
C^m$  without diagonals)  and its universal enveloping algebra $UX^m$ is
prounipotent completion of the group algrbra or pure braid group (see [K2] and
references therein). Kohno [K1] used this algebra in order to write
the most general form of Knizhnik-Zamolodchikov equation [KZ]

$$ {{d \Psi }\over {dz_i}} = \hbar \sum_{j \neq i} {{t^{ij}} \over
{z_i-z_j}}\Psi    \eqno    (2.7)$$

where  $\psi$  is $UX^m$ -valued meromorphic function on  $\bf (C^m$ without
diagonals),
$\hbar = {{h }\over {2 \pi i}}$
Relations (2.6) are imposed in order to preserve zero-curvature
condition
$$[{{d} \over {dz_i}} - \hbar \sum_{j\neq i} {{t^{ij}} \over {z_i-z_j}}     ;
{{d}\over {dz_k}} - \hbar \sum_{l\neq k} {{t^{kl}} \over {z_k-z_l}} ] = 0
\eqno  (2.8)$$
which allows us to construct monodromy representation of pure braid
group in  the group $exp(X^m) \i UX^m$ . This representation is nonlocal and
its  matrix elements are certain multiple integrals .
     Following Drinfel'd [D1] let us put m=3 and write
differential equation
$$ {{dG(x)}\over {dx}} = \hbar ({{t^{12}}\over {x}} + {{t^{23}}\over
{x-1}})G(x) \in UX^3   \eqno  (2.9)$$

Than any solution  $\Psi (z_1, z_2, z_3)$ of  KZ-equation (2.7) with m=3
has the form
 $$ {(z_1-z_3)^{\hbar (t^{12}+t^{13}+t^{23})}}G{{(z_1-z_2)}  \over {(z_1-z_3)}}
\eqno (2.10)$$
where  G  satisfies (2.9)
     Let  $G_1$  and  $G_2$  be solutions of (2.9) defined when
$0<x<1$  with the asymptotic behavior  $$G_1(x) \approx  x^{\hbar t^{12}}  (x
\rightarrow 0)$$
and   $$G_2(x) \approx (x-1)^{\hbar t^{23}}  (x \rightarrow 1)$$.
Then         $$ G_1 = G_2 \Phi_{kz}   \eqno (2.11)$$
for some   $\Phi_{kz} \in exp(X^3) \i UX^3$   and
$$ W_1 = W_2 \Phi_{kz}   \eqno  (2.12)$$
where $W_1$  and  $W_2$  are solutions of (2.7) with m=3 defined
in the domain  $z_1 > z_2 > z_3$ with the asymptotic behavior
$$W_1 \approx (z_1-z_2)^{\hbar t^{12}} (z_1-z_3)^{\hbar (t^{13}+t^{23}}$$
  for $(z_1-z_2) << (z_1-z_3)$  and
$$W_2 \approx (z_2-z_3)^{\hbar t^{23}} (z_1-z_3)^{\hbar (t^{13}+t^{12}}$$
        for $(z_2-z_3) << (z_1-z_3)$ .

	As it was proved in [Dr1], [Dr2]
 $$ \Phi_{kz} = \phi_{kz}(\hbar t^{12}, \hbar t^{23})  \eqno(2.13)$$
, where $\phi_ {kz}(A, B)$ is some
element of graded completion of tensor algebra with two generators A and B;
 $log(\phi_{kz}(A, B)) = P_{kz}(A, B)$   belongs to the graded completion of
free Lie
algebra withtwo generators A and B; and
$$ P_{kz}(B, A) = -P_{kz}(A, B)  \eqno (2.14)$$

     Algebra  $UX^m$   can be imbedded in the algebra  $A_{kz}^m$
of Feynman diagrams (see [BN1],[BN5])of the form depicted on fig 6 .

\setlength{\unitlength}{0.0125in}%
\begin{picture}(175,180)(130,515)
\thicklines
\multiput(250,630)(8.46154,0.00000){7}{\line( 1, 0){  4.231}}
\multiput(200,565)(20.26667,27.02222){3}{\line( 3, 4){  8.667}}
\multiput(250,630)(-7.77778,0.00000){14}{\line(-1, 0){  3.889}}
\multiput(145,600)(8.46154,0.00000){7}{\line( 1, 0){  4.231}}
\put(305,535){\vector( 0, 1){160}}
\put(200,535){\vector( 0, 1){160}}
\put(145,535){\vector( 0, 1){160}}
\put(130,515){\makebox(0,0)[lb]{\raisebox{0pt}[0pt][0pt]{\elvrm m  upward
pointed arrows}}}
\put(235,540){\makebox(0,0)[lb]{\raisebox{0pt}[0pt][0pt]{\elvrm . . .}}}
\end{picture}

\centerline  {\bf                 {fig.6}}

These diagrams are defined in the same way as usual BN-diagrams,
but they have  m  upward pointed Wilson lines instead of one Wilson
loop. Here  $t^{ij}$    is presented by  BN-diagram on fig.7.

\setlength{\unitlength}{0.0125in}%
\begin{picture}(270,205)(85,505)
\thicklines
\multiput(180,615)(7.82609,0.00000){12}{\line( 1, 0){  3.913}}
\put(355,515){\vector( 0, 1){195}}
\put(270,520){\vector( 0, 1){190}}
\put(180,515){\vector( 0, 1){195}}
\put( 90,515){\vector( 0, 1){195}}
\put(295,530){\makebox(0,0)[lb]{\raisebox{0pt}[0pt][0pt]{\elvrm . . .}}}
\put(200,530){\makebox(0,0)[lb]{\raisebox{0pt}[0pt][0pt]{\elvrm . . .}}}
\put(105,530){\makebox(0,0)[lb]{\raisebox{0pt}[0pt][0pt]{\elvrm . . .}}}
\put(350,505){\makebox(0,0)[lb]{\raisebox{0pt}[0pt][0pt]{\elvrm m}}}
\put(265,510){\makebox(0,0)[lb]{\raisebox{0pt}[0pt][0pt]{\elvrm j}}}
\put(175,505){\makebox(0,0)[lb]{\raisebox{0pt}[0pt][0pt]{\elvrm i}}}
\put( 85,505){\makebox(0,0)[lb]{\raisebox{0pt}[0pt][0pt]{\elvrm 1}}}
\end{picture}

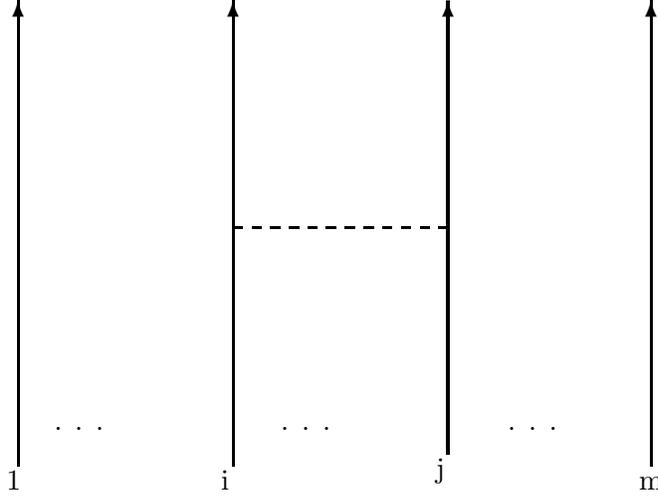
\centerline {\bf                 {fig.7}}

and the diagram with  2n vertices (this number is always even) is
said to be if degree  n. The multiplication in this algebra of
diagrams is just putting one diagram over another. It is easy to
see that above defined grading and multiplication in  $A_{kz}^m$    are
compatible with those of  $UX^m$ .

     We'll need for ous purposes to define a semi-direct product
$Y^m$  of the group algebra  $\bf C S_m$   of symmetric group and  $A_{kz}^m$
as follows: Y   is generated as a linear space by pairs  (x, s);
x  is diagram from $A_{kz}^m$ '  $s \in S_m$   with multiplication
$$(x_1 , s_1 )(x_2 , s_2 )=(x_1 s(x_2 ), s_1s_2 )$$ Here we suppose that
symmetric
group   acts on  $A_{kz}^m$  by permutations of strings.  Algebra $Y^m$ has an
important subgroup  $$G^m = S_m * exp(X^m) \i Y^m$$
     Let  $s_i  (1 \le i \le m-1)$ be the standard generators of the braid
group $B_m$  (if it not lead to confusion, we'll denote the elementary
transpositions $s_i \in S_m$ by the same symbols.

     Let us define representation  $\rho$: $B_m \rightarrow G^m \i Y^m$  as
follows:

$$ \rho (s_{m-1}) = (e^{ht^{m-1, m}\over 2} ; s_{m-1} )  \eqno  (2.15)$$

$$ \rho(s_i) = \phi_{kz}^{-1}(ht^{i, i+1};h\sum_{s=i+2}^m t^{i+1, s})
(e^{ht^{i, i+1}\over 2} ; s_i)\phi_{kz}(ht^{i, i+1};h\sum_{s=i+2}^m t^{i+1, s})
  \eqno (2.16)$$

     if $1 \le i \le m-1$

This construction of representation $\rho$  is due to Drinfeld [Dr2]. It may be
called
"the universal" braid group representation since the group  $G^m$ can be
interpreted
as prounipitent completion of $B_m$ [Dr2], [K2].

	Above constructed representation $\rho $ is generalization of braid group
action on quasitensor category [Dr1],[Dr2],[Re1]. To construct representation
of the braid group $B_m$ one has to choose some configuration of parentheses in
(nonassociative) product of  m  symbols  $x_1,...,x_m$. Then, in order to
define action of the braid group generator  $s_i$, we should:

a) change the configuration of papentheses in order to have
$...(x_ix_{i+1})...$ inside one pair of parantheses (this gives us some
transition operator $\Phi_{trans}$,constructed out of "associator" $\Phi_{kz}$
according to Drinfeld's rules [Dr1],[Dr2]),

b) apply the Drinfeld's R-matrix $(e^{ht^{i, i+1}\over 2} ; s_i)$ ,and

c) return back to our initial configuration of parantheses (this gives us
inverse operator to the operator $\Phi_{trans}$  .

	Formulas (2.15) and (2.16) correspond to one particular choice of
configuration of papentheses, namely, $(x_1(x_2(...(x_{m-1}x_m))...)$ but any
other choice is possible as well and gives us equivalent representation with
some transition operator $\Phi_{trans}$ as an intertwinier. (If it don't lead
to confusion, we'll denote all "transition operators" corresponding to
transitions between different configurations of parentheses, by the same symbol
 $\Phi_{trans}$).
$$ $$

\centerline {\bf   { 3.  TAKING THE TRACE}}
\centerline { }
     It is well-known (see, for instance,  [Bi] that any oriented s-component
link  L  can be presented as a closed braid. Two braids $b_1 \in B_{m1}$    and
$b_2 \in B_{m2}$ give under closure the same link iff they can be obtained
from each other by finite sequence of Markov moves of two types:
 $$ b_1b_2 \approx b_2b_1 \in B_m  \eqno (3.1)$$                $$ b \in B_m
\approx  bs_m^{\pm 1} \in B_{m+1}    \eqno   (3.2)$$
    Thus,  any function  $f :\bigcup_m {B_m} \rightarrow F^s $ for some
commutative
ring  K  and some  K-module $F^s$  gives rise to some link
invariant iff  f takes the same values on braids,  equivalent
with respect to (3.1) and (3.2). Any framed link also can be
presented as a closed braid. The analogues of Markov moves for
braids which give under closure the same framed link (with
blackboard framing [Tu], [Pi3], [Pi4] ) can also be described
explicitly (see [Re2 ]). Here we  give sufficient conditions for
function $f :\bigcup_m {B_m}\rightarrow F^s$  to descend to some framed link
invariant:
 $$ f(b_1b_2) = f(b_2b_1) ,    (b_1 ;b_2 \in  B_m) \eqno         (3.1A)$$
 $$ f(bs_m^{\pm 1} ) = q_i^{\pm 1}  f(b);   b \in B_m \eqno
(3.2A)$$
  for some invertible element  $q_i \in K$ (here  i  is number of the link
component ).
    In the case of knots,  let us put  $F = \oplus_n F_n$  to be (graded)
completion of Kontsevich Hopf algebra,  considered as a module over
graded completion of $\bf C [F_1]$. Here $q = e^{ht\over 2}$  ,  where  t is
the standard generator in  $F_1$ (see fig.8)
\setlength{\unitlength}{0.0125in}%
\begin{picture}(262,172)(124,529)
\thicklines
\put(255,615){\oval(262,172)}
\multiput(125,615)(8.00000,0.00000){33}{\line( 1, 0){  4.000}}
\end{picture}

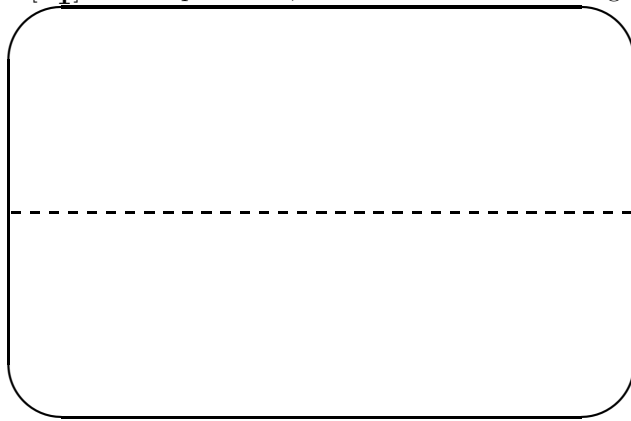
\centerline { \bf                      {fig.8} }

      If we want to
consider invariants of s-component links,  we should take $F^s$  to
be the $\bf C$ -module  $F^s$,  generated by BN-diagrams with s Wilson loops .
Here we put  $q_i = e^{ht_i\over 2}$ .
	 Let us fix configuration of papentheses in (nonassociative) product of 2m
symbols  $x_1,...,x_m,y_m,...,y_1$ as follows :
$$((x_1((x_2(...((x_{m-1}(x_my_m))y_{m-1})...))y_2))y_1)   \eqno (3.3)$$
Then, using this configuration of parantheses,let us define, in the spirit of
formulas (2.15),(2.16) and remarks after them, representation $\widehat
{\rho}$: $B_m \rightarrow G^{2m} \i Y^{2m}$ as restriction on $B_m \i B_{2m}$
of representation  $\Phi_{trans}^{-1} \rho \Phi_{trans}$:$B_{2m} \rightarrow
G^{2m} \i Y^{2m}$ , where $\Phi_{trans}$ is transition operator between "the
standard" configuration of parentheses on the set of 2n elements and the
configuration (3.3).

	Now let us suppose that the first m Wilson lines in $Y^{2m}$ are oriented "up"
and the second m wilson lines are oriented "down".Then for any  $m \i \bf N$
let us
define map

$\tau : Y^{2m} \rightarrow \bigcup_{s=1}^m F^s$ of graded linear spaces ,which
is defined
simply by closure of  2m  directed Wilson lines in one Wilson  loop
according to the rule,  given by underlying permutation  $s \in \bf {S_m}$
(see
fig.9 as an example)

\setlength{\unitlength}{0.0125in}%
\begin{picture}(400,233)(5,584)
\thicklines
\put(155,760){\vector( 0,-1){150}}
\put(180,760){\vector( 0,-1){145}}
\put( 85,610){\vector( 0, 1){155}}
\put(115,760){\vector( 0,-1){150}}
\multiput( 85,695)(8.57143,0.00000){4}{\line( 1, 0){  4.286}}
\put( 45,610){\vector( 0, 1){150}}
\put( 10,615){\vector( 0, 1){150}}
\multiput( 10,665)(7.89474,0.00000){10}{\line( 1, 0){  3.947}}
\multiput( 45,620)(12.91137,32.27842){2}{\line( 2, 5){  5.365}}
\multiput( 10,640)(7.77778,0.00000){5}{\line( 1, 0){  3.889}}
\multiput( 85,640)(8.23529,0.00000){9}{\line( 1, 0){  4.118}}
\put(150,605){\makebox(0,0)[lb]{\raisebox{0pt}[0pt][0pt]{\elvrm 2}}}
\put(175,605){\makebox(0,0)[lb]{\raisebox{0pt}[0pt][0pt]{\elvrm 1}}}
\put(150,760){\makebox(0,0)[lb]{\raisebox{0pt}[0pt][0pt]{\elvrm 2}}}
\put(175,760){\makebox(0,0)[lb]{\raisebox{0pt}[0pt][0pt]{\elvrm 1}}}
\put( 80,605){\makebox(0,0)[lb]{\raisebox{0pt}[0pt][0pt]{\elvrm 3}}}
\put(110,605){\makebox(0,0)[lb]{\raisebox{0pt}[0pt][0pt]{\elvrm 3}}}
\put( 75,765){\makebox(0,0)[lb]{\raisebox{0pt}[0pt][0pt]{\elvrm 1}}}
\put(110,765){\makebox(0,0)[lb]{\raisebox{0pt}[0pt][0pt]{\elvrm 3}}}
\put( 40,760){\makebox(0,0)[lb]{\raisebox{0pt}[0pt][0pt]{\elvrm 3}}}
\put(  5,765){\makebox(0,0)[lb]{\raisebox{0pt}[0pt][0pt]{\elvrm 2}}}
\put( 40,605){\makebox(0,0)[lb]{\raisebox{0pt}[0pt][0pt]{\elvrm 2}}}
\put(  5,610){\makebox(0,0)[lb]{\raisebox{0pt}[0pt][0pt]{\elvrm 1}}}
\put(270,610){\vector( 0, 1){0}}
\put(320,610){\oval(100, 20)[bl]}
\put(320,615){\oval(120, 30)[br]}
\put(310,610){\vector( 0, 1){0}}
\put(325,610){\oval( 30, 10)[bl]}
\put(325,610){\oval( 30, 10)[br]}
\put(335,755){\vector( 0,-1){0}}
\put(303,755){\oval( 64, 52)[tr]}
\put(303,755){\oval( 66, 52)[tl]}
\put(375,760){\vector( 0,-1){0}}
\put(305,760){\oval(140,112)[tr]}
\put(305,760){\oval(140,112)[tl]}
\put(400,755){\vector( 0,-1){0}}
\put(353,755){\oval( 94, 66)[tr]}
\put(353,765){\oval( 86, 46)[tl]}
\put(235,615){\vector( 0, 1){0}}
\put(317,615){\oval(164, 60)[bl]}
\put(317,620){\oval(176, 70)[br]}
\put(310,610){\vector( 0, 1){155}}
\put(378,758){\vector( 0,-1){150}}
\put(233,613){\vector( 0, 1){150}}
\put(268,608){\vector( 0, 1){150}}
\put(338,758){\vector( 0,-1){150}}
\put(201,680){\vector( 1, 0){ 20}}
\multiput(308,638)(8.23529,0.00000){9}{\line( 1, 0){  4.118}}
\multiput(233,638)(7.77778,0.00000){5}{\line( 1, 0){  3.889}}
\multiput(268,618)(12.91137,32.27842){2}{\line( 2, 5){  5.365}}
\multiput(233,663)(7.89474,0.00000){10}{\line( 1, 0){  3.947}}
\multiput(308,693)(8.57143,0.00000){4}{\line( 1, 0){  4.286}}
\put(403,758){\vector( 0,-1){145}}
\end{picture}

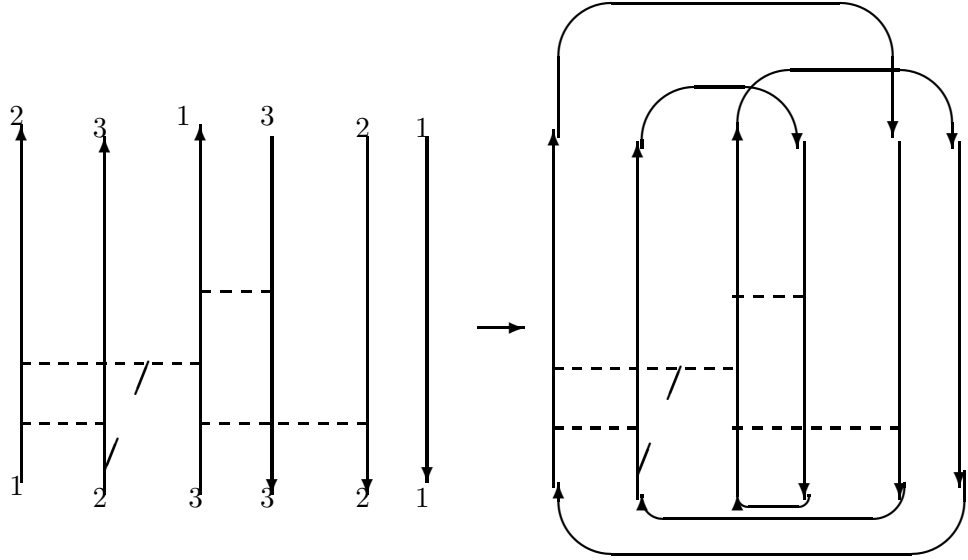
 \centerline { \bf  {      fig.9}}

if it will not lead to confusion, we'll not distingoish braids in $B_m$ and
their images in $Y^{2m}$. We will also suppose number of components  s  to pe
fixed, omit index m  and write $\tau :\bigcup_m B_m \rightarrow  F^s$

 Let  $b_1 \in B_{m_1}$ ; $b_2 \in B_{m_2}$   be two braids,
$b_2$ gives knot under closure,  and let $(b_1*b_2) \in  B_{m_1+m_2-1}$
be the braid,  obtained from  $b_1$  and  $b_2$  by the procedure shown on
fig.10.

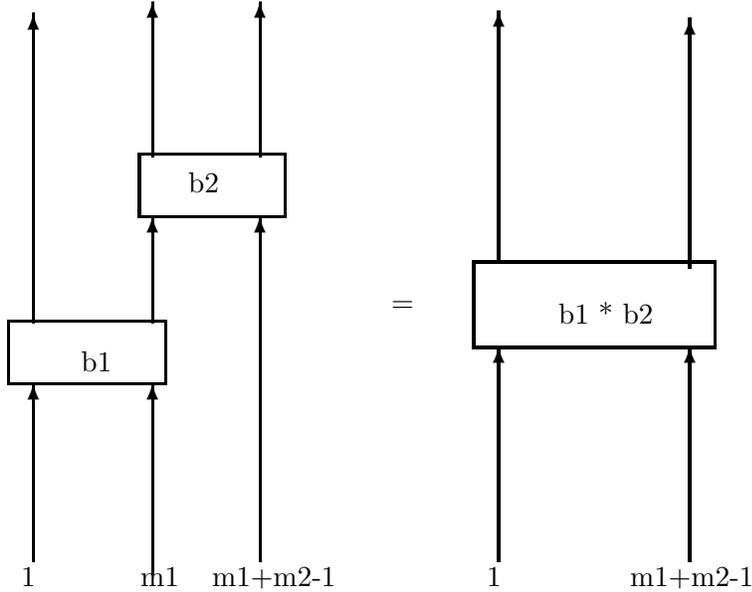
\setlength{\unitlength}{0.0125in}%
\begin{picture}(295,245)(55,470)
\thicklines
\put(250,570){\framebox(100,35){}}
\put(285,580){\makebox(0,0)[lb]{\raisebox{0pt}[0pt][0pt]{\elvrm b1 * b2}}}
\put(110,625){\framebox(60,25){}}
\put(130,635){\makebox(0,0)[lb]{\raisebox{0pt}[0pt][0pt]{\elvrm b2}}}
\put(340,480){\vector( 0, 1){ 90}}
\put(260,480){\vector( 0, 1){ 90}}
\put(160,480){\vector( 0, 1){145}}
\put(115,475){\vector( 0, 1){ 80}}
\put( 65,580){\vector( 0, 1){130}}
\put( 65,480){\vector( 0, 1){ 75}}
\put( 55,555){\framebox(65,25){}}
\put(340,603){\vector( 0, 1){105}}
\put(260,606){\vector( 0, 1){105}}
\put(115,580){\vector( 0, 1){ 45}}
\put(115,650){\vector( 0, 1){ 65}}
\put(160,650){\vector( 0, 1){ 65}}
\put(215,585){\makebox(0,0)[lb]{\raisebox{0pt}[0pt][0pt]{\elvrm =}}}
\put(315,470){\makebox(0,0)[lb]{\raisebox{0pt}[0pt][0pt]{\elvrm m1+m2-1}}}
\put(255,470){\makebox(0,0)[lb]{\raisebox{0pt}[0pt][0pt]{\elvrm 1}}}
\put( 60,470){\makebox(0,0)[lb]{\raisebox{0pt}[0pt][0pt]{\elvrm 1}}}
\put(110,470){\makebox(0,0)[lb]{\raisebox{0pt}[0pt][0pt]{\elvrm m1}}}
\put(140,470){\makebox(0,0)[lb]{\raisebox{0pt}[0pt][0pt]{\elvrm m1+m2-1}}}
\put( 85,560){\makebox(0,0)[lb]{\raisebox{0pt}[0pt][0pt]{\elvrm b1}}}
\end{picture}

\centerline {\bf    {fig.10}}

\proclaim Theorem 3.1.
$$\tau (b_1 * b_2) = \tau (b_2) * \tau (b_1) $$ , where * be the action of $F$
on $F^s$ (on the s-th component).

	Proof. geometrically obvious from  (3.3),fig.9 and
fig.10.
\centerline { }

	Let $q = e^{ht\over 2}$ and let $\mu \in F$ be the
image of Drinfeld's associator $\Phi_{kz} \in exp(X^3) \i UX^3$ under "the
closure map" shown on fig.11.

\setlength{\unitlength}{0.0125in}%
\begin{picture}(426,258)(10,482)
\thicklines
\put( 10,605){\framebox(145,30){}}
\put( 25,635){\line( 0, 1){100}}
\put( 80,635){\line( 0, 1){ 95}}
\put(135,635){\line( 0, 1){ 95}}
\put( 25,605){\line( 0,-1){ 80}}
\put( 80,605){\line( 0,-1){ 85}}
\put(135,605){\line( 0,-1){ 90}}
\put( 35,610){\makebox(0,0)[lb]{\raisebox{0pt}[0pt][0pt]{\elvrm associator}}}
\put(312,530){\oval(194, 94)[bl]}
\put(312,606){\oval(246,246)[br]}
\put(345,606){\oval(180,238)[tr]}
\put(245,727){\oval( 60, 22)[tr]}
\put(245,727){\oval( 60, 22)[tl]}
\put(319,527){\oval( 88, 30)[bl]}
\put(319,517){\oval( 52, 10)[br]}
\put(165,620){\vector( 1, 0){ 20}}
\put(200,612){\framebox(165,30){}}
\put(215,642){\line( 0, 1){ 85}}
\put(275,642){\line( 0, 1){ 85}}
\put(275,727){\line(-1, 0){  5}}
\put(215,612){\line( 0,-1){ 85}}
\put(275,612){\line( 0,-1){ 85}}
\put(345,642){\line( 0, 1){ 85}}
\put(345,612){\line( 0,-1){ 90}}
\put(235,622){\makebox(0,0)[lb]{\raisebox{0pt}[0pt][0pt]{\elvrm associator}}}
\end{picture}

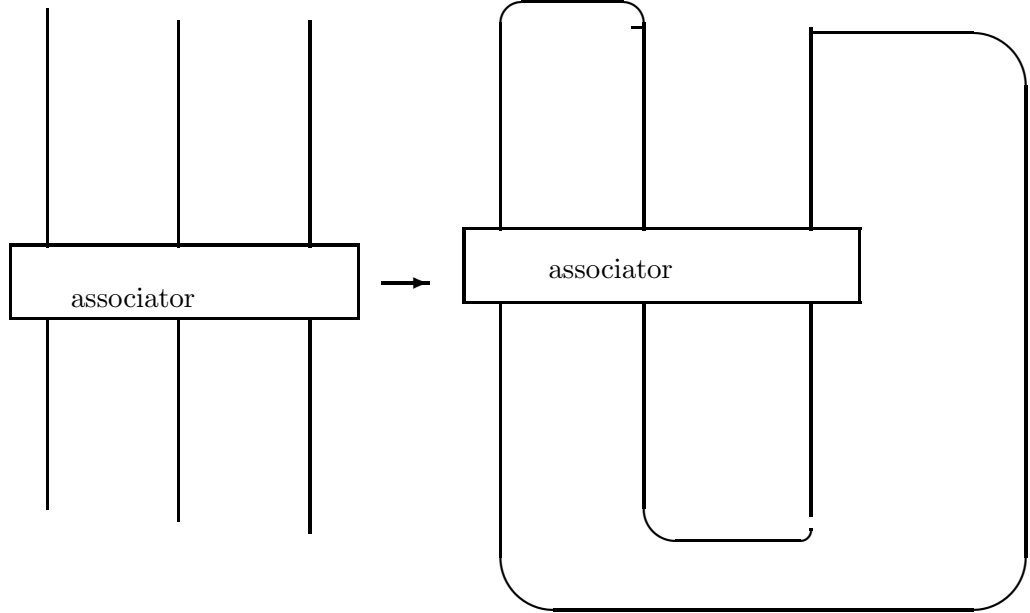
 \centerline { \bf  {      fig.11}}

\proclaim  Remark. Above defined $\mu $ is equal to the value of generating
function of Kontsevich integrals on the Morse
knot shown on fig.12.

\setlength{\unitlength}{0.0125in}%
\begin{picture}(85,132)(60,569)
\thicklines
\put(105,635){\oval( 30, 12)[bl]}
\put(105,635){\oval( 30, 12)[br]}
\put(103,585){\oval( 86, 30)[bl]}
\put(103,585){\oval( 84, 30)[br]}
\put(133,660){\oval( 24, 10)[tr]}
\put(133,660){\oval( 26, 10)[tl]}
\put( 75,695){\oval( 30, 10)[tr]}
\put( 75,695){\oval( 30, 10)[tl]}
\put(145,660){\line( 0,-1){ 75}}
\put(145,655){\makebox(0.4444,0.6667){\tenrm .}}
\put(120,660){\line( 0,-1){ 25}}
\put( 90,695){\line( 0,-1){ 60}}
\put( 90,690){\makebox(0.4444,0.6667){\tenrm .}}
\put( 60,585){\line( 0, 1){110}}
\end{picture}

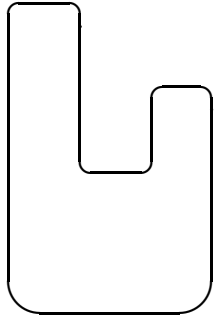
 \centerline { \bf  {      fig.12}}

\proclaim     Lemma 3.2. Let $s_1$ be the standard generator of $B_2$. Then
$\tau (s_1^{\pm 1}) = q^{\pm 1}\mu $

We give here pictorial proof:

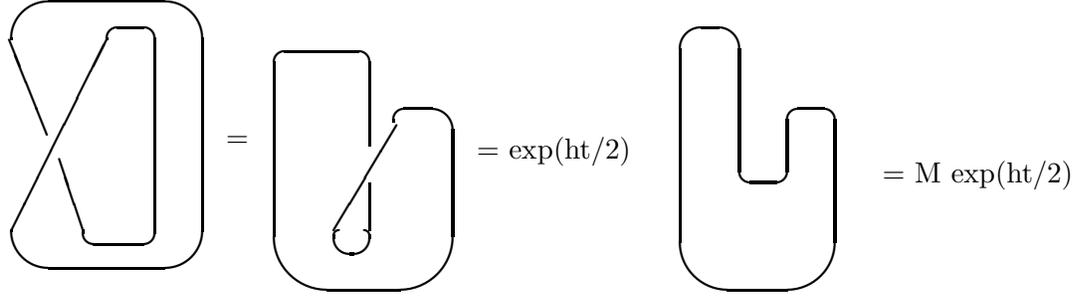
\setlength{\unitlength}{0.0125in}%
\begin{picture}(365,123)(35,604)
\thicklines
\put(348,625){\oval( 66, 40)[bl]}
\put(348,625){\oval( 64, 40)[br]}
\put(328,705){\oval( 24, 20)[tr]}
\put(328,705){\oval( 26, 20)[tl]}
\put(370,675){\oval( 20, 12)[tr]}
\put(370,675){\oval( 20, 12)[tl]}
\put(350,655){\oval( 20, 10)[bl]}
\put(350,655){\oval( 20, 10)[br]}
\put(170,628){\oval(  0,  4)[tl]}
\put(178,628){\oval( 16, 16)[bl]}
\put(178,628){\oval( 14, 16)[br]}
\put(185,628){\oval(  0,  4)[tr]}
\put(183,630){\oval( 76, 50)[bl]}
\put(183,630){\oval( 74, 50)[br]}
\put(206,670){\oval( 28, 22)[tr]}
\put(206,675){\oval( 22, 12)[tl]}
\put(165,700){\oval( 40, 10)[tr]}
\put(165,700){\oval( 40, 10)[tl]}
\put( 75,630){\oval( 80, 32)[bl]}
\put( 75,630){\oval( 80, 32)[br]}
\put( 80,630){\oval( 30, 12)[bl]}
\put( 80,630){\oval( 30, 12)[br]}
\put( 75,710){\oval( 80, 32)[tr]}
\put( 75,710){\oval( 80, 32)[tl]}
\put( 85,710){\oval( 20, 10)[tr]}
\put( 85,710){\oval( 20, 10)[tl]}
\put(380,675){\line( 0,-1){ 50}}
\put(360,655){\line( 0, 1){ 20}}
\put(340,705){\line( 0,-1){ 50}}
\put(315,625){\line( 0, 1){ 80}}
\put(220,670){\line( 0,-1){ 40}}
\put(185,665){\line( 0, 1){ 35}}
\put(185,630){\line( 0, 1){ 20}}
\put(170,630){\line( 3, 5){ 26.471}}
\put(145,630){\line( 0, 1){ 70}}
\put(115,710){\line( 0,-1){ 80}}
\put( 95,710){\line( 0,-1){ 80}}
\put( 50,670){\line(-2, 5){ 15.862}}
\put( 65,630){\line(-1, 3){ 10}}
\put( 35,630){\line( 1, 2){ 40}}
\put(400,650){\makebox(0,0)[lb]{\raisebox{0pt}[0pt][0pt]{\elvrm =   M
exp(ht/2)}}}
\put(230,660){\makebox(0,0)[lb]{\raisebox{0pt}[0pt][0pt]{\elvrm = exp(ht/2)}}}
\put(125,665){\makebox(0,0)[lb]{\raisebox{0pt}[0pt][0pt]{\elvrm =}}}
\end{picture}

 \centerline { \bf  {      fig.13}}
The first identity in fig.13 follows from identity on fig.14

\setlength{\unitlength}{0.0125in}%
\begin{picture}(473,375)(5,435)
\thicklines
\put( 98,525){\oval( 14, 10)[tr]}
\put( 98,525){\oval( 16, 10)[tl]}
\put(223,525){\oval(  4,  0)[tr]}
\put(223,520){\oval( 26, 10)[tl]}
\put(228,745){\oval(  4,  0)[tr]}
\put(228,740){\oval( 26, 10)[tl]}
\put(133,745){\oval( 14, 10)[tr]}
\put(133,745){\oval( 16, 10)[tl]}
\put(260,708){\oval( 60,206)[br]}
\put(260,708){\oval( 60,204)[tr]}
\put(145,710){\oval( 70,190)[tl]}
\put(145,710){\oval( 70,190)[bl]}
\put( 55,760){\oval(  4, 10)[br]}
\put( 50,760){\oval( 14, 14)[tr]}
\put( 50,765){\oval( 10,  4)[tl]}
\put(336,710){\oval( 60,190)[tl]}
\put(336,710){\oval( 60,190)[bl]}
\put(373,701){\oval( 50,172)[br]}
\put(363,701){\oval( 70,198)[tr]}
\put(331,750){\oval( 14, 10)[tr]}
\put(331,750){\oval( 16, 10)[tl]}
\put(441,745){\oval( 14, 10)[tr]}
\put(441,745){\oval( 16, 10)[tl]}
\put(105,525){\vector( 0,-1){ 90}}
\put( 90,435){\vector( 0, 1){ 95}}
\multiput( 90,480)(7.64706,0.00000){9}{\line( 1, 0){  3.824}}
\put(155,435){\vector( 0, 1){100}}
\multiput(225,485)(7.27273,0.00000){6}{\line( 1, 0){  3.636}}
\put(265,435){\vector( 0, 1){ 95}}
\put(210,440){\vector( 0, 1){ 85}}
\put(225,525){\vector( 0,-1){ 85}}
\put(230,745){\vector( 0,-1){ 85}}
\put(215,660){\vector( 0, 1){ 85}}
\put(270,655){\vector( 0, 1){ 95}}
\multiput(230,705)(7.27273,0.00000){6}{\line( 1, 0){  3.636}}
\put(140,745){\vector( 0,-1){ 90}}
\put(125,655){\vector( 0, 1){ 95}}
\multiput(125,700)(7.64706,0.00000){9}{\line( 1, 0){  3.824}}
\put(190,655){\vector( 0, 1){100}}
\put( 45,675){\vector(-1, 4){  8.823}}
\put( 25,735){\vector(-1, 4){  8.823}}
\put(  5,685){\vector( 1, 2){ 40}}
\put( 55,755){\vector(-1,-2){ 38}}
\put(338,750){\vector( 0,-1){ 90}}
\put(323,660){\vector( 0, 1){ 95}}
\put(368,657){\vector( 0, 1){100}}
\put(478,649){\vector( 0, 1){100}}
\put(433,655){\vector( 0, 1){ 95}}
\put(448,745){\vector( 0,-1){ 90}}
\put( 15,485){\makebox(0,0)[lb]{\raisebox{0pt}[0pt][0pt]{\elvrm since}}}
\put(290,480){\makebox(0,0)[lb]{\raisebox{0pt}[0pt][0pt]{\elvrm = 0}}}
\put(175,480){\makebox(0,0)[lb]{\raisebox{0pt}[0pt][0pt]{\elvrm +}}}
\put(297,705){\makebox(0,0)[lb]{\raisebox{0pt}[0pt][0pt]{\elvrm *}}}
\put( 45,720){\makebox(0,0)[lb]{\raisebox{0pt}[0pt][0pt]{\elvrm = exp(h/2}}}
\put(200,700){\makebox(0,0)[lb]{\raisebox{0pt}[0pt][0pt]{\elvrm +}}}
\put(411,710){\makebox(0,0)[lb]{\raisebox{0pt}[0pt][0pt]{\elvrm =}}}
\end{picture}

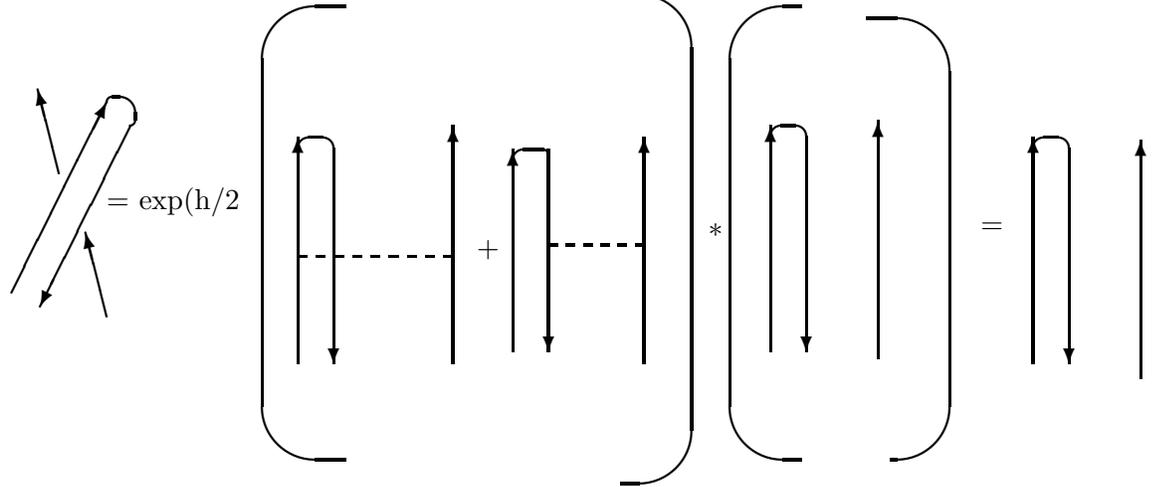
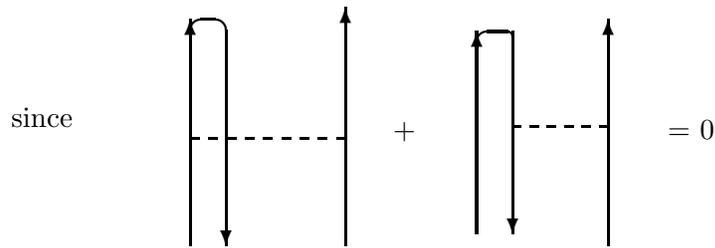
 \centerline { \bf  {      fig.14}}
Let $\bf J: B_m \rightarrow \bigcup_s F^s$ be
equal to $(\mu)^{1-m} \tau: B_m \rightarrow \bigcup_s F^s $.

\proclaim Lemma 3.3.
 The map  $\bf J$  is "Markov trace" i.e.,  it satisfies
    (3.1A) and (3.2A).

    Proof.

a) the property (3.1A) is geometrically obvious.

b) the property (3.2A) follows from theorem 3.1 and lemma 3.2.
from which (3.2A) with   $q = e^{ht\over 2}$   can easily be seen. The lemma is
proved.

     Let $\bf J$  be above defined framed link invariant.
     Let us consider its perturbative expansion:$J = \sum_{n=0}^{\infty}  h^n
J_n$

\proclaim     Lemma 3.4.  $J_n$  is  $F_n^s$ - valued Vassiliev framed link
invariant of order n.
Proof:

     Since for any braid $ b \in B_m$  if
$\widehat {\rho }(b) = \sum_{n=0}^{\infty} x_n(b) h^n \i   Y^{2m}$  ,
that $ x_n (b) \i Y^{2m}$   has degree  n  in  $Y^{2m}$   (this fact is true
for the
generators  $s_i \in B_m$  and thus,  for any  $b \in B_m$),  then for any
framed oriented link  $\bf L$  $J_n(\bf L)$ also has degree  n. Thus
$J_n(\bf L) \in F_n^s \i F^s$.
     Let  $\bf L$  be singular imbedding of  $(\bf {S^1})^s$   to  $\bf {R^3}$
 with (n+1) double crossing
points. Than it can be presented as a closure of "generalized
braid" [Pi1], [Ba] (braid where in some places generatots  $s_i$
are changed on the generators  $a_i$ ,  depicted on fig.15.

 \centerline { }
\setlength{\unitlength}{0.0125in}%
\begin{picture}(290,160)(65,590)
\thicklines
\put(355,600){\vector( 0, 1){145}}
\put(275,600){\vector( 0, 1){150}}
\put(235,600){\vector(-1, 3){ 50.500}}
\put(175,600){\vector( 1, 2){ 74}}
\put(145,600){\vector( 0, 1){150}}
\put( 70,605){\vector( 0, 1){145}}
\put(295,620){\makebox(0,0)[lb]{\raisebox{0pt}[0pt][0pt]{\elvrm . . .}}}
\put( 85,620){\makebox(0,0)[lb]{\raisebox{0pt}[0pt][0pt]{\elvrm . . .}}}
\put(225,590){\makebox(0,0)[lb]{\raisebox{0pt}[0pt][0pt]{\elvrm i+1}}}
\put(170,595){\makebox(0,0)[lb]{\raisebox{0pt}[0pt][0pt]{\elvrm i}}}
\put( 65,595){\makebox(0,0)[lb]{\raisebox{0pt}[0pt][0pt]{\elvrm 1}}}
\end{picture}

                          $$ \eqno           (3.4)  $$

 \centerline {\bf  {fig.15}}
\centerline { }
     The representation  $\rho : B_m \rightarrow Y^m$   can be extended to
these
"generalized braids" by formula
 $$\rho(a_i) = \rho (s_i ) - \rho (s_i^{-1}) \eqno (3.5) $$  (and the
representation  $\widehat {\rho }: B_m \rightarrow Y^{2m}$  can be extended to
"generalized braids" by the same formula).

	(3.5), (2.14) and (2.15) imply that
$$\rho(a_{m-1}) = {2sh{ht^{m-1, m}\over 2}; s_{m-1}} \eqno   (3.6) $$
$$ \rho(a_i) = \phi_{kz}^{-1}(ht^{i, i+1};h\sum_{s=i+2}^m t^{i+1, s})
(2sh{ht^{i, i+1}\over 2} ; s_i)\phi_{kz}(ht^{i, i+1};h\sum_{s=i+2}^m t^{i+1,
s})  \eqno (3.7)$$

                                           if $i<m-1$

     Thus  $\widehat {\rho }(a_i)$  are divisible by  h in  $Y^{2m} \otimes \bf
{C} {[}h]$,  which implies,
that for any "generalized braid"  $b \in B_m$  with (n+1) double
crossing points  $\widehat {\rho }(b)$  is divisible by
 $h^{n+1}$   . Thus
 $\bf J(\bf L)$  is
divisible by  $h^{n+1}$   ,  which means that  $J_n(\bf L)$ = 0 for any
singular
embedding  $\bf L$  with (n+1) double crossing points,  i.e., $J_n$ is
Vassiliev invariant of order  n . Lemma is proved.

     Let  $V_n^s$ be the space of Vassiliev invariants of  framed s-component
links of
order  n. Then there is a natural map   $f_n$  : $V_n^s \rightarrow
V_n^s/V_{n-1}^s \rightarrow (F_n^s)^* $,
defined as follows [Va1], [Va2], [BL].  Let  g  be some Vassiliev  invariant of
order  n and  D be Vassiliev [n]-diagram. Then
    $$ (J_n(g);D)  = (g;L(D))    \eqno       (3.8)$$
where  L(D) be some singular embedding  $(\bf {S^1})^s \rightarrow \bf {R^3}$
with  n  double
crossing points and the underlying configuration of  n  points on
$(\bf {S^1})^s$   given by diagram  D.
\proclaim     Theorem 3.5.   The map ($J_n$;...):$(F_n^s)^* \rightarrow V_n^s$
 is  left inverse
to $f_n$,  and differs from its right inverse on Vassiliev invariant
of order  n-1.

     Proof.
It is sufficient to prove that for any singular embedding
L: $(\bf {S^1})^s \rightarrow \bf R^3$   with precisely  n  double points
    $$ J_n(L)=D(L)   \eqno     (3.9)$$
where D(L)  is Vassiliev  [n]-diagram with  n propagators,  joining
those points on  $(\bf {S^1})^s $ ,  which are identified under  L. Let us
present
L  as a closure of some "generalized braid"  $b \in  B_m$ . Then  $\widehat
{\rho }(b)$
is product of some terms of the form  $$ (e^{ht^{i, i+1}\over 2} ;  s_{i} )
\eqno (3.10)$$

 $$\Phi_{trans}^{\pm 1} \eqno (3.11)$$
 $$ (2sh{ht^{i, i+1}\over 2}; s_{i})  \eqno (3.12) $$
There are precisely  n  terms of forms of form (3.12).
Since terms (3.10) and (3.11) have the form
 $${1+hX}   \eqno   (3.13)$$
for some $X \in Y^{2m}$ ; $\mu^{\pm 1}$ also has the form (3.13) for some $X
\in F$ ; and the terms (3.12) have the form
 $$ ht^{i,i+1} + h^2 X     \eqno    (3.14)$$
for some $X \in Y^{2m}$ , then the expression for the coefficient in  $h^n$  in
perturbative expansion of  $\widehat {\rho }(b)$
has only one term,  which gives  $J_n$(L)=D(L) , as desired.
The theorem is proved.

     Kontsevich Hopf algebra F has (graded) factor algebra
$A = F/F_1F$. Then  $A_n^*$   is canonically identified with
the space of Vassiliev unframed knot invariants of order  n
factored through the space of invariants of order  n-1. In the
basis of Vassiliev [n]-diagrams in  $F_n$ the projector
P:  $F_n \rightarrow A_n$  can be described explicitly  [Pi2].
$$P(D) = \sum_{k=0}^{n} (-t)^n \sum_I D_I  \eqno    (3.15)$$
Where  t  is the generator of  $F_1$; the second sum in (3.15) is taken over
all  [k]-subdiagrams $D_I$ of  D.
     The quantity $ P(J) = \sum_{n=0}^{\infty}  h^n P(J_n)$ :
{Knots}$\rightarrow $  F  is the
generating function for "universal" (order n)-Vassiliev knot
invariants and has the same formal properties than the generating function $I =
\sum_{n=0}^{\infty}  h^n I_n$  of Kontsevich integrals [Ko1].

\proclaim      Theorem 3.6.  Let  $K_1$  and  $K_2$  be two oriented framed
knots;  $K_1*K_2$  be their connected sum. Then $J(K_1*K_2) = J(K_1) J(K_2)$.

      Proof. It follows immediately from theotem 3.1 and the definition of  J .

$$ $$

 \centerline {\bf       {4. DISCUSSIONS}}

     At the moment there are three different expressions for the
universal Vassiliev knot invariant,  (the quantity,  which satisfies
conditions of theorems 3.3 and 3.6). The first one is
constructed from perturbative expansion of monodromy of KZ-equation
(Kontsevich integrals [Ko1]),  the second one is constructed from
perturbative Chern-Simons theory [Ko1], [BN3] (These integrals are highly
singular. Their  convergence is rather difficult to prove). The third
construction is presented here(see also [Ko5] where the similiar combinatorial
construction was given, using knot diagram and a point on it). This
construction is the best one
for computations.

     The "universal Vassiliev invariant" in the form presented here
can be evaluated purely combinatorially for any particular link
L ,  if we know explicit expression for an "associator" $\Phi_{kz} =
\phi_{kz}(\hbar t^{12}, \hbar t^{23}) \in X^3$   as
a formal noncommutative power series in $\hbar t^{12}$ and   $\hbar t^{23}$ .
The powerful algebraic techniques to find this expression for $\Phi_{kz}$ was
developed in  [Dr2].An "iterated integral" expression for "assosiator" was
proposed in
[BN6], which proves immediately equivalence of our approach with Kontsevich's
one (see also [LM] where coincidence of coefficients $P_{sl_m,V_{fund},n}$ of
one-variable reductions of HOMFLY polynomial  and Kontsevich integrals $I_n$,
evaluated on the weight system $C_{sl_m,V_{fund},n} \in F_n^*$ is proved).

	The analogous problem for "Kontsevich integrals"
is much more complicated and involves calculations with hypergeomertic type
integrals [TK].In our approach only one such integral (for each  $n \in {\bf
N}$) should be calculated. The calculations in perturbative Chern-Simons
theory are even more complicated and are hardly to be accomplished
by direct methods.

	Above defined construction $\tau $ of  the universal Vassiliev invariant of
link which can be presented as a closure of braid  has a straightforvard
gneralization on arbitrarylink diagram, and even on string link diagram
[BN5].Roughly speaking,  $\tau $ is decomposition of the generating function of
Kontsevich integrals (berore inserting the correction factor $\mu^{1-m}$) in
the product of "elemrntary" factors corresponding to decomposition of link
diagram on "elementary" pieces.

	Drinfeld`s construction of representation  $\rho$: $B_m \rightarrow G^m$
involving $\Phi_{kz} = \phi_{kz}(\hbar t^{12}, \hbar t^{23})$ in its definition
uses only the following properties of  $\phi_{kz}$:
$$ \phi_{kz}(\hbar t^{12}, \hbar (t^{23}+t^{24})) \phi_{kz}(\hbar
(t^{12}+t^{13}), \hbar t^{34}) = $$
$$= \phi_{kz}(\hbar t^{23}, \hbar t^{34}) \phi_{kz}(\hbar (t^{12}+t^{13}),
\hbar (t^{24}+t^{34})) \phi_{kz}(\hbar t^{12}, \hbar t^{23}) \in exp(X^4)
\eqno         (5.1)$$
$$e^{ht^{12}+ht^{13}\over 2} = \Phi ^{312}e^{ht^{13}\over 2}(\Phi ^{132})^{-1}
e^{ht^{23}\over 2}\Phi \in exp(X^3)  \eqno (5.2)$$
    $$e^{ht^{13}+ht^{23}\over 2} = (\Phi ^{231})^{-1}e^{ht^{13}\over 2}\Phi
^{213}
e^{ht^{12}\over 2}(\Phi )^{-1}\in exp(X^3)  \eqno (5.3)$$
$$\Phi \in X^3  \eqno
  (5.4)$$
	Here $\Phi ^s$ ($s \in S_3$) is the image of $\Phi \in UX^3$ under
automorphism  $s: t^{ij} \rightarrow t^{s(i)s(j)}$
	Thus, if we find any other solution of (5.1)-(5.4) we can also construct
"the universal braid group representation" $\rho_{\phi}$ and "the universal
Vassiliev knot invariant" $J^{\phi}$. If we put $\phi$ to be formal
noncommutative power series
with rational coefficients,  we obtain the "universal $\bf Q$-valued Vassiliev
invariant (the quantity,  which satisfies conditions of theorems 3.3 and 3.6
and the condition, that for any oriented s-component link  L
 $$J_n^{\phi}(L) \i F_n^s\otimes \bf Q  \i F_n^s \otimes \bf C$$ (here we use
the definition of $F_n$ as an abelian group i.e., over $\bf Z$).

     To understand better all these formally different
"universal formulas" and make calculations with them one needs
some unification theorem,  stating,  that all these quantities  (with different
$\phi $)
either coincide or are some functions of each other. We conjecture
existence of such a theorem.
\centerline {  }
\centerline {\bf  Acknowledgements}
     I benefited much from discussions with D.Bar-Natan, V.Drinfeld, D.Kazhdan
and expecially with M.Kontsevich to whom I express my sincere gratitude.

\centerline {\bf Literature}

\noindent [AS]   S.Axelrod,  I.Singer. Chern-Simons perturbation theory.

       MIT preprint,  October 1991.

\noindent [Ar]   V.I.Arnold. Private communication. April 1992.

\noindent [Ba]   J.C.Baez,  Link invariants of finite type and perturbation
theory.

Wellesley college preprint,  July 1992.

\noindent [Bi]   J.S.Birman,   Braids,  links and mapping class groups.

       Ann.of Math.Studies, no 82, Princeton Univ.Press.
       Princeton,  NJ, 1974.

\noindent [BL]   J.S.Birman and X.-S.Lin,  Knot polynomials and Vassiliev's
       invariants. Preprint.1991.

\noindent [BN1]  D.Bar-Natan,  On Vassiliev knot invariants. Preprint.
       August 1992.

\noindent [BN2]  D.Bar-Natan,  Weights of Feynmann diagrams and Vassiliev

       knot invariants.
 Preprint.February 1991.

\noindent [BN3]  D.Bar-Natan,  Perturbative aspects of the Chern-Simons
       topological quantum field theory.Ph.D.thesis, Princeton
       Univ., June 1991, Dep.of Mathematics.

\noindent [BN4]  D.Bar-Natan,  Perturbative Chern-Simons theory. Preprint.
       August 1990.

\noindent [BN5] D.Bar-Natan, Vassiliev homotopy string link invariants.

Preprint.December 1992.

\noindent [BN6] D.Bar-Natan, Private communication.

\noindent [Co]   J.H.Conway,  An enumeration of Knots and Links and some
       of their Algebraic Properties. in Computational Problems
       Abstract Algebra, p.329-358,

 New York, Pergamon Press, 1969.

\noindent [Dr1]  V.G.Drinfeld,  Quasi-Hopf algebras,

Leningrad  Math J.,vol.1
       (1990)p.1419-1457.

\noindent [Dr2]  On quasitriangular quasi-Hopf algebras and a group,  closely
connected

 with Gal(Q/Q). Leningrad Math J.,vol.2 (1991)p.829-860.

\noindent [Dr3]  V.G.Drinfeld,  Quantum groups, Proc. Intern. Congr.Math.

       Berkeley, 1(1986),  p.798-820.

\noindent [GMM]  E.Guadagnini,  M.Marellini and M.Mintchev,

  Perturbative
       aspects of the Chern-Simons field theory. Phys.Lett.
       B227(1989)p.111

\noindent [HOMFLY]P.Freyd,  D.Yetter,  J.Hoste,  W.B.R.Lickorish,  K.Milett,
       A.Ocneanu.

A new polynomial invariant of knots and
       links.

Bull. Amer. Math., Soc.(1985), v.12, No2, p.239-246.

\noindent [Jo1]  V.F.R.Jones,  A polynomial invariant for links via von
       Neumann algebras. Bull. AMS, v.12(1985), p.103-110.

\noindent [Jo2]  V.F.R.Jones,  On knot invariants related to  some
       statistical mechanical models. Pacific  J.Math.
       v.137 (1989)p.311-334.

\noindent [K1]   T.Kohno. Monodromy representations of braid groups and
classical Yang-Baxter                       c      equations, Ann.Inst.Fourier,
 Grenoble, v.37(1987), p.139-160.

\noindent [K2]   T.Kohno. Series de Poincare-Koszul associe aux grouppes de
tresses pures. Inv.Math. No 82(1985), p.57-75

\noindent [Ka1]  L.H.Kauffman,  An invariant of regular isotopy.
       Trans.Amer.Math.Soc., vol.318, 2 (1990) 317-371.

\noindent [Ka2]  L.H.Kauffman,  State model and the Jones polynomial.
       Topology,  v.26(1987)p.395-407.

\noindent [Ko1]  M.Kontsevich,  Vassiliev's knot invariants. Preprint 1992 .

\noindent [Ko2]  M.Kontsevich,  Private communication.

\noindent [Ko3]  M.Kontsevich, Feynman diagrams and Low-dimensional topology.

 Preprint, January 1993.

\noindent [Ko4]  M.Kontsevich, Graphs, homotopical algebra and low-dimensional
topology.

 Pre-preprint, 1992.

\noindent [Ko4]  M.Kontsevich, in preparation.

\noindent [KZ]   V.G.Knizhnik and A.B.Zamolodchikov,  Current algebra and
    Wess-Zumino models in two dimensions,  Nucl.Phys.B247
       (1984)p.83-103.

\noindent [Li1]  X.S.Lin,  Vertex models, quantum group and Vassiliev knot
       invariants.  Preprint,  1991

\noindent [Li2]  X.S.Lin,  Milnor $\mu $-invarianys ar all of finite type.
       Preprint,  1992.

\noindent [Mi1]  J.Milnor, Link groups. Ann.of Math.59(1954), p.177-195.

\noindent [LM] Le Tu Quoc Thang and J.Murakami. On relation between
Kontsevich's integral invariant and invariants coming from quantum R-matrices.

    Pre-preprint.January 1993.

\noindent [Mi2]  J.Milnor,  Isotopy of links,  Algebraic geometry and Topology
       (R.H.Fox ed.) Princeton Univ.Press., Princeton,  No1,  (1957)
       p.280-306.

\noindent [Pi1]  S.A.Piunikhin,  Weights of Feynmann diagrams,  link
       polynomials and Vassiliev knot invariants. To appear
       in Journal of Knot Theory and its Ramifications .

\noindent [Pi2]  S.A.Piunikhin, Vassiliev knot invariants contain more
information,  than all
       knot polynomials (To appear in Journal of Knot Theory and its
       Ramifications)

\noindent [Pi3] S.A.Piunikhin, Turaev-Viro and Kauffman-Lins Invariants for
3-Manifolds
     coincide. Journal of Knot Theory and its Ramifications,

vol 1, No2
       (1992), p.105-135.

\noindent [Pi4]  S.A.Piunikhin,   State sum model for trivalent knotted graph
       invariants using the quantum group SL (2).

 Journal of Knot Theory and its Ramifications , v.1, No3(1992), p.273-278.

\noindent [Ro]   D.Rolfsen,  Knot and Links,  (Publish or Perish, 1976).

\noindent [Re1]  N.Yu.Reshetikhin,   Quantized universal enveloping algebras,
       the Yang-Baxter equation and invariants of links.
       LOMI-preprint E-4-87,  E-17-87.

\noindent [Re2]  N.Yu.Reshetikhin,   Quasitriangular Hopf algebras and
       invariants of links. Leningrad,  Math J.1 (1990)p.

\noindent [RT]   N.Yu.Reshetikhin and V.G.Turaev,  Ribbon graphs and
       their invariants derived from quantum groups.
       Commun.Math.Phys., vol.127, 1(1990) 1-26.

\noindent [TK]   A.Tsuchiya, Y.Kanie Vertex operators in two dimensional
conformal field theory on $\bf P^1$ and monodromy representations of braid
groups.

 in Advanced studies in Pure Mathematics. 16. Tokyo, Kinokuniya, 1988,
p.297-372.

\noindent [Tu]   V.G.Turaev,  The Yang-Baxter equation and invariants of
       links.

Invent.Math.v.92(1988)p.527-553.

\noindent [Tu]   V.G.Turaev,  Operator invariants of tangles and R-matrices,
       Math.UssR,  Izvestiya,  vol.35., (1990), No2, p.411-444.

\noindent [Va1]  V.A.Vassiliev,  Cohomology of Knot Spaces. Advances in
       Soviet Mathematics,  v.1, AMS(1990), p.23-69 .

\noindent [Va2]  V.A.Vassiliev,  Complements of discriminants of smooth maps.
       Topology and applications,  Trans.of Math.Mono.98, Amer.Math.
       Soc., Providence,  1992.

\noindent [Wi]   E.Witten,  Quantum field theory and Jones polynomial.

       Commun.Math.Phys., vol.121, No3 (1989) p.351-399.

$$ $$

\end{document}